\documentclass[11pt]{llncs}
\usepackage{preamble}

\begin{document}
\title{A Puff of Steem: Security Analysis of Decentralized Content Curation}
\author{Aggelos Kiayias\inst{1,2} \and Benjamin Livshits\inst{3,4} \and Andr\'es
Monteoliva Mosteiro\inst{1,5} \and Orfeas Stefanos Thyfronitis Litos\inst{1}}
\institute{University of Edinburgh \and IOHK \and Imperial College of London
\and Brave Software \and Clearmatics \\
\email{akiayias@inf.ed.ac.uk}, \email{ben@brave.com},
\email{amonteolivam@gmail.com}, \email{o.thyfronitis@ed.ac.uk}}
\maketitle

\thispagestyle{plain}
\begin{abstract}
  Decentralized content curation is the process through which uploaded posts are
  ranked and filtered based exclusively on users' feedback. Platforms such as
  the blockchain-based Steemit\footnote{\url{https://steemit.com/} Accessed:
  2019-01-02} employ this type of curation while providing monetary incentives
  to promote the visibility of high quality posts according to the perception of
  the participants. Despite the wide adoption of the platform very little is
  known regarding its performance and resilience characteristics. In this work,
  we provide a formal model for decentralized content curation that identifies
  salient complexity and game-theoretic measures of performance and resilience
  to selfish participants. Armed with our model, we provide a first analysis of
  Steemit identifying the conditions under which the system can be expected to
  correctly converge to curation while we demonstrate its susceptibility to
  selfish participant behaviour. We validate our theoretical results with system
  simulations in various scenarios.
\end{abstract}

\section{Introduction}
  The modern Internet contains an immense amount of data; a single user can only consume a tiny fraction in a reasonable amount of time. Therefore, any widely used platform that hosts user-generated content (UGC) must employ a content curation mechanism.
   Content curation can be understood as the set of mechanisms which rank, aggregate and filter relevant information. In recent years, popular news aggregation sites like Reddit\footnote{\url{https://www.reddit.com/} Accessed: 2019-01-02} or Hacker News\footnote{\url{https://news.ycombinator.com/} Accessed: 2019-01-02} have established crowdsourced curation as the primary way to filter content for their users.
   Crowdsourced content curation, as opposed to more traditional techniques such as expert- or algorithmic-based curation, orders and filters content based on the ratings and feedback of the users themselves, obviating the need for a central moderator by leveraging the ``wisdom of the crowd''~\cite{askalidis2013theoretical}.

  The decentralized nature of crowdsourced curation makes it a suitable solution for ranking user-generated content in blockchain-based content hosting systems. The aggregation and filtering of user-generated content emerges as a particularly challenging problem in permissionless blockchains, as any solution that requires a concrete moderator implies that there exists a privileged party, which is incompatible with a permissionless blockchain.
   Moreover, public blockchains are easy targets for Sybil attacks, as any user can create new accounts at any time for a marginal cost.
    Therefore, on-chain mechanisms to resist the effect of Sybil users are necessary for a healthy and well-functioning platform; traditional counter-Sybil mechanisms~\cite{levine2006survey} are much harder to apply in the case of blockchains due to the decentralized nature of the latter.
   The functions performed by moderators in traditional content platforms need to be replaced by incentive mechanisms that ensure self-regulation. Having the impact of a vote depend on the number of coins the voter holds is an intuitively appealing strategy to achieve a proper alignment of incentives for users in decentralized content platforms; specifically, it can render Sybil attacks impossible.

   However, the correct design of such systems is still an unsolved problem. Blockchains have created a new economic paradigm where users are at the same time equity holders in the system, and leveraging this property in a robust manner constitutes an interesting challenge.
   A variety of projects have designed decentralized content curation systems~\cite{synereo,steemit,tcr}. Nevertheless, a deep understanding of the properties of such systems is still lacking. Among them, Steemit
has a long track record, having been in operation since 2016 and attaining a user base of more than 1.08 M\footnote{\url{https://steemdb.com/accounts} Accessed: 2019-01-02} registered accounts\footnote{The number of accounts should not be understood as the number of active users, as one user can create multiple accounts.}. Steemit is a social media platform which lets users earn money (in the form of the STEEM cryptocurrency) by both creating and curating content in the network. Steemit is the front-end of the social network, a graphical web interface which allows users to see the content of the platform. On the other hand, all the back-end information is stored on a distributed ledger, the Steem blockchain. Steem can be understood as an ``app-chain'', a blockchain with a specific application purpose: serving as a distributed database for social media applications~\cite{steemit}.

\noindent  \textbf{Our Contributions.}
In this work we study the foundations of decentralized content curation from a computational perspective. We develop an abstract model of a post-voting system which aims to sort the posts created by users in a distributed and crowdsourced manner.
  Our model is constituted by a functionality which executes a protocol performed by $N$ players. The model includes an honest participant behaviour while it allows deviations to be modeled for a subset of the participants.      The $N$ players contribute votes in a round-based curation process. The impact of each vote depends on the number of coins held by the player.
   The posts are arranged in a list, sorted by the value of votes received, resembling the front-page model of Reddit or Hacker News. In the model, players vote according to their subjective opinion on the quality of the posts and have a limited attention span.

   Following previous related work~\cite{ghosh2011incentivizing,askalidis2013theoretical}, we represent each player's opinion on each post (i.e. likability) with a numerical value $\like \in [ 0,1 ]$.
   The objective quality of a post is calculated as the simple summation of all players' likabilities for the post in question. 
   To measure the effectiveness of a post-voting system, we introduce the property of  \textit{convergence} under honesty which is parameterised by a number of values including a metric $t$, that demands the  first $t$ articles to be   ordered according to the objective quality of the posts at the end of the execution assuming all participants signal honestly to the system their personal preferences. 
Armed with our  post-voting system abstraction, we proceed to particularize it to model Steemit and provide the following results. 

\begin{itemize}
\item[i)] We characterise the conditions under which the Steemit algorithm converges under honesty. Our results highlight some fundamental limitations of the actual Steemit parameterization. Specifically, for curated lists of length bigger than 70 the algorithm may {\em not achieve even 1-convergence}. 
\item[ii)]
We validate our results with a simulation testing different metrics based on correlation that have been proposed in previous works~\cite{kendall1955rank,spearman1904proof} and relating them to our notion of convergence.
\item[iii)]
We demonstrate that ``selfish'' deviation from honest behavior results to substantial gains in terms of boosting the ranking of specific posts in the resulting list of the post-voting system.
\end{itemize}






\section{Related Work}
User-generated content (UGC) has been identified as a fundamental component of social media platforms and Web 2.0 in general~\cite{kaplan2010users}. The content created by users needs to be curated, and crowdsourced content curation~\cite{askalidis2013theoretical} has emerged as an alternative to expert-based~\cite{stanoevska2012content} or algorithmic-based~\cite{rader2015understanding} curation techniques. Motivated by the widespread adoption of crowdsourced aggregation sites such as Reddit or Digg\footnote{\url{http://digg.com/} Accessed: 2019-01-02}, several research efforts~\cite{das2010ranking,ghosh2011incentivizing,abbassi2014distributed} have aimed to model the mechanics and incentives for users in UGC platforms. This surge of interest is accompanied by studies which have shown how
social media users behave strategically when they publish and consume content~\cite{may2014filter}. As an example, in the case of Reddit, users try to maximize their `karma'~\cite{bergstrom2011don}, the social badge of the social media platform~\cite{anderson2013steering}.

Previous works have analyzed content curation from an incentives and game-theoretic standpoint~\cite{ghosh2011incentivizing,das2010ranking,gupte2009news,may2014filter,abbassi2014distributed} . Our formalisation is based on these models and inherits features such as the quality distribution of the articles and the users' attention span~\cite{askalidis2013theoretical,ghosh2011incentivizing}. In terms of the analysis of our results, the analysis of our \textit{$t$-convergence} metric is similar to the top-$k$ posts in~\cite{askalidis2013theoretical}. We also leverage the rank correlation coefficients Kendall's Tau~\cite{kendall1955rank} and Spearman's Rho~\cite{spearman1904proof} to measure content curation efficiency.
Our approach describes the mechanics of post-voting systems from a computational perspective, something that departs from the approach of all previous works, drawing inspiration from the real-ideal world paradigm of cryptography~\cite{goldreich1999foundations,lindell} as employed in our definition of $t$-convergence.

Post-voting systems constitute a special case of voting mechanisms, as studied within social choice theory, belonging to the subcategory of cardinal voting systems~\cite{hillinger2005case}. In this context, it follows from Gibbard's theorem~\cite{gibbard1973manipulation} that no decentralised non-trivial
post-voting mechanism can be strategy-proof. This is consistent with our results that
demonstrate how selfish behaviour is beneficial to the participants. Our system shares the property of spanning multiple voting rounds with previous work~\cite{kalech2011practical}. Other related literature in social choice~\cite{lu2011robust,conitzer2005communication,xia2010compilation} is centered on political elections and as a result attempts to resolve a variation of the problem with quite different constraints and assumptions. In more detail, in the case of political elections, voter communication in many rounds is costly while navigating the ballot is not subject to any constraints as voters are assumed to have plenty of time to parse all the options available to them. As a result, voters can express their preferences for any candidate, irrespective of the order in which the latter appear on the ballot paper. On the other hand, the online and interactive nature of post-voting systems make multi-round voting a natural feature to be taken advantage of. At the same time, the fairness requirements are more lax and it is acceptable (even desirable) for participants to act reactively on the outcome of each others' evaluations. On the other hand, in the post-voting case, the ``ballot'' is only partially available given the high number of posts to be ranked that may very well exceed the time available to a (human) user to participate in the process. As a result a user will be unable to vote for posts that she has not viewed, for instance, because they are placed in the bottom of the list. This is captured in our model by introducing the concept of ``attention span.''

Content curation is also related to the concept of online governance. The governance of online communities such as Wikipedia has been thoroughly studied in previous academic work~\cite{leskovec2010governance,forte2008scaling}. However, the financially incentivized governance processes in blockchain systems, where the voters are at the same time equity-holders, have still many open research questions~\cite{vitalik,ehrsam}. This shared ownership property has triggered interest in building social media platforms backed by distributed ledgers, where users are rewarded for generated content and variants of coin-holder voting are used to decide how these rewards are distributed.
The effects of explicit financial incentives on the quality of content in Steemit has been analyzed in~\cite{thelwall2017can}.
Beyond the Steemit's whitepaper~\cite{steemit}, a series of blog posts~\cite{curationRewards,selfvoters} effectively extend the economic analysis of the system. In parallel with Steemit, other projects such as Synereo~\cite{synereo} and Akasha\footnote{\url{https://akasha.world/} Accessed: 2019-01-02} are exploring the convergence of social media and decentralized content curation.
Beyond blockchain-based social media platforms, coin-holder voting systems are present in decentralized platforms such as DAOs~\cite{darkdaos} and in different blockchain protocols~\cite{dash,tezos}. However, most of these systems use coin-holder voting processes to agree on a value or take a consensual decision.

\section{Model}
  We first introduce some useful notation:
\begin{itemize}
  \item We denote an ordered list of elements with $A = \left[e_1, \dots,
  e_n\right]$ and the $i$-th element of the list with $A\left[i\right] = e_i$.
  \item Let $n \in \mathbb{N}^*$. $\left[n\right]$ denotes $\left\{1, 2, \dots,
  n\right\}$.
\end{itemize}

  \subsection{Post list}
    \begin{definition}[Post]
  Let $\playerlen \in \mathbb{N}^*$. A post is defined as $\post = \left(m,
  \like\right)$, with $m \in \left[\playerlen\right], \like \in \left[0,
  1\right]^\playerlen$.
  \begin{itemize}
    \item \textbf{Author.} The first element of a post is the id of its creator,
    $m$.

    \item \textbf{Likability.} The likability of a post is defined as $\like \in
    \left[0, 1\right]^\playerlen$.
  \end{itemize}
\end{definition}

    \noindent $\playerlen$ represents the number of voters (a.k.a. players). A
    post has a distinct likability in $\left[0, 1\right]$ for each player.

    \begin{definition}[Ideal Score of a post]
  Let post $\post = \left(m, \like\right)$. We define the \emph{ideal score} of
  $\post$ as $\idsc{\post} = \sum\limits_{i = 1}^{|\like|} \like_i$.
\end{definition}

    \noindent The ideal score of a post is a single number that represents its
    overall worth to the community. By using simple summation, we assume that
    the opinions of all players have the same weight.

    \begin{definition}[Post List]
  Let $\postlen \in \mathbb{N}^*$. A post list $\postlist = \left[\post_1,
  \dots, \post_\postlen\right]$ is an ordered list containing posts. It may be
  the case that two posts are identical.
\end{definition}

    \noindent In the case of many UGC platforms, e.g. Steemit, there exists a
    feed (commonly named ``Trending'') that displays the same ordered posts for
    all users. In such an ordered list, posts placed closer to the top are more
    visible, since users typically consume content from top to bottom. We can
    thus measure the quality of an ordered list of posts by comparing it with a
    list that contains the same posts in decreasing order of ideal score.

    \begin{definition}[$t$-Ideal Post Order]
  Let $\postlist$ a list of posts, $t \in [M]$. The property
  $\textsc{Ideal}^{t}\left(\postlist\right)$ holds if
  \begin{equation*}
    \forall i < j \in [t], \idsc{\postlist\left[i\right]} \geq
    \idsc{\postlist\left[j\right]} \enspace.
  \end{equation*}
  We say that $\postlist$ has a $t$-\emph{ideal rank} if
 $\textsc{Ideal}^{t}\left(\postlist\right)$ holds and $t$ is the maximum integer less or
  equal to $M$ with this property.

\end{definition}

  \subsection{Post Voting System}

    We now define an abstract post-voting system. Such a system is defined
    through two Interactive Turing Machines (ITMs),
    $\mathcal{G}_{\mathrm{Feed}}$ and $\Pi_{\mathrm{honest}}$. The first
    controls the list of posts and aggregates votes, whereas one copy of the
    second ITM is instantiated for each player. $\mathcal{G}_{\mathrm{Feed}}$
    sends the post list to one player at a time, receives her vote and reorders
    the post list accordingly. The process is possibly repeated for many rounds.

    A measure of the quality of a post-voting system is the $t$-ideal rank of
    the post list at the end of the process.

    In a more general setting, some of the honest protocol instantiations may be
    replaced with an arbitrary ITM. A robust post-voting system should still
    produce a post list of high quality.

    \begin{definition}[Post-Voting System]
  \label{model:def:pvs}
  Consider four PPT algorithms $\textsc{Init}, \textsc{Aux},
  \textsc{HandleVote}$ and $\textsc{Vote}$. The tuple $\mathcal{S}$ consisting
  of the four algorithms is a Post-Voting System. $\mathcal{S}$ parametrizes the
  following two ITMs:

  $\gfunc$ is a global functionality that accepts two messages: \texttt{read},
  which responds with the current list of posts and \texttt{vote}, which can
  take various arguments and does whatever is defined in \textsc{HandleVote}.

  $\honeststr$ is a protocol that sends \texttt{read} and \texttt{vote}
  messages to $\gfunc$ whenever it receives (\texttt{activate}) from $\env$.
\end{definition}
\begin{algorithm}[H]
  \caption{$\gfunc\left(\textsc{Init}, \textsc{Aux},
  \textsc{HandleVote}\right)\left(\postlist, \mathrm{initArgs}\right)$}
  \label{alg:gfunc}
  \begin{algorithmic}[1]
    \State Initialization:
    \Indent
      \State $\playerlist \gets \emptyset$ \Comment{Set of players}
      \State $\textsc{Init}\left(\mathrm{initArgs}\right)$
    \EndIndent
    \State
    \State Upon receiving (\texttt{read}) from $\player_{\pid}$:
    \Indent
      \State $\playerlist \gets \playerlist \cup \left\{\player_{\pid}\right\}$
      \State $\mathrm{aux} \gets \textsc{Aux}\left(\player_{\pid}\right)$
      \State Send (\texttt{posts}, $\postlist$, aux) to $\player_{\pid}$
    \EndIndent
    \State
    \State Upon receiving (\texttt{vote}, ballot) from
    $\player_{\mathrm{pid}}$:
     \Indent
       \State \textsc{HandleVote}(ballot)
     \EndIndent
  \end{algorithmic}
\end{algorithm}

\begin{algorithm}[H]
  \caption{$\honeststr\left(\textsc{Vote}\right)$}
  \label{alg:honest}
  \begin{algorithmic}[1]
    \State Upon receiving (\texttt{activate}) from $\env$:
    \label{alg:honest:activate}
    \Indent
      \State Send (\texttt{read}) to $\gfunc$
      \State Wait for response (\texttt{posts}, $\postlist$, aux)
      \State $\mathrm{ballot} \gets \textsc{Vote}\left(\postlist,
      \mathrm{aux}\right)$
      \State Send (\texttt{vote}, ballot) to $\gfunc$
    \EndIndent
  \end{algorithmic}
\end{algorithm}

    \noindent Players are activated by an Environment ITM that sends activation
    messages (Algorithm~\ref{alg:honest}, line~\ref{alg:honest:activate}).

    \begin{definition}[Post-Voting System Activation Message]
  We define $\mathtt{act}_\pid$ as the message $\left(\mathtt{activate},
  \pid\right)$, sent to $\player_\pid$.
\end{definition}

    \begin{definition}[Execution Pattern]
  \label{model:def:execpat}
  Let $\playerlen, \rounds \in \mathbb{N}^*, \playerlen \geq 2$.
  \begin{gather*}
    \execpat_{\playerlen, \rounds} = \left\{\left(\mathtt{act}_{\pid_1}, \dots,
    \mathtt{act}_{\pid_{\playerlen \rounds}}\right) :\right. \\
    \left.\forall i \in \left[\rounds\right], \forall k \in
    \left[\playerlen\right], \exists j \in \left[\playerlen\right]: \pid_{(i -
    1) \playerlen + j} = k\right\} \enspace,
  \end{gather*}
  i.e. activation messages are grouped in $\rounds$ rounds and within each round
  each player is \texttt{activate}d exactly once. The order of activations is
  not fixed.

  Let Environment $\env$ that sends messages $\mathrm{msgs} =
  \left(\mathtt{act}_{\pid_1}, \dots, \mathtt{act}_{\pid_n}\right)$
  sequentially. We say that \emph{$\env$ respects $\execpat_{\playerlen,
  \rounds}$} if $\mathrm{msgs} \in \execpat_{\playerlen, \rounds}$. (Note: this
  implies that $n = \playerlen \rounds$.)
\end{definition}

    \begin{definition}[$\left(\playerlen, \rounds, \postlen, t\right)$-convergence
under honesty]
  We say that a post-voting system $\pvs = \left(\textsc{Init}, \textsc{Aux},
  \textsc{HandleVote}, \textsc{Vote}\right)$ $\left(\playerlen, \rounds,
  \postlen, t\right)$-\emph{converges under honesty} (or \emph{$t$-converges
  under honesty for $\playerlen$ players, $\rounds$ rounds and $\postlen$
  posts}) if, for every input $\postlist$ such that $|\postlist| = \postlen$,
  for every $\env$ that respects $\execpat_{\playerlen, \rounds}$ and given that
  all protocols execute $\honeststr$, it holds that after $\env$ completes its
  execution pattern, $\gfunc$ contains a post list $\postlist'$ such that
  $\textsc{Ideal}^t\left(\postlist'\right)$ is true.
\end{definition}

    \noindent Note that concrete post voting systems may or may not give
    information such as the total number of rounds $\rounds$ to the players.
    This is decided in algorithm \textsc{Aux}.

    We now give a high-level description of a concrete post voting system, based
    on the Steemit platform. According to this mechanism, each player is
    assigned a number of coins known as ``Steem Power'' ($\stpow$) that remains
    constant throughout the execution and another number called ``Voting Power''
    ($\votpow$) in $\left[0, 1\right]$, initialized to 1. A vote is a pair
    containing a post and a weight $w \in \left[0, 1\right]$. Upon receiving a
    list of posts, the honest player chooses to vote her most liked post amongst
    the top $\attspan$ posts of the list. The weight is chosen to be equal to
    the respective likability. The functionality increases the score of the post
    by $\stpow\left(a \votpow w + b\right)$ and subsequently decreases the
    player's Voting Power by the same amount (but keeping it within the
    aforementioned bounds).

    \begin{definition}[Steemit system]
  The Steemit system is the post voting system $\pvs$ with parameters $a, b,
  \regen \in \left[0, 1\right]: a + b < 1, \ceil*{\frac{a + b}{\regen}} > 1,
  \attspan \in \mathbb{N}^*, \stpowvec \in \mathbb{R}_{+}^\playerlen$. The four
  parametrizing procedures can be found in Appendix~\ref{appendix:procs}.
\end{definition}

    \begin{remark}
  The constraint $a + b < 1$ ensures that a single vote of full weight cast by a
  player with full Voting Power does not completely deplete her Voting Power.
  The constraint $\ceil*{\frac{a + b}{\regen}} > 1$ excludes the degenerate
  case in which the regeneration of a single round is enough to fully replenish
  the Voting Power in all cases; in this case the purpose of Voting Power would
  be defeated.
\end{remark}

\begin{remark}
  \label{remark:steemitparms}
  The Steem blockchain protocol defines $a = 0.02, b = 0.0001$ and $\regen =
  \frac{3}{5 \cdot 24 \cdot 60 \cdot 60} = 0.0000069\bar{4}$, thus
  $\ceil*{\frac{a + b}{\regen}} = 2895$. A post can be voted for 7 days from its
  creation and at most one vote can be cast every 3 seconds, thus $\rounds =
  \frac{7 \cdot 24 \cdot 60 \cdot 60}{3} = 201600$.
\end{remark}

\begin{remark}
  Note (Algorithm~\ref{alg:steem:vote},
  lines~\ref{alg:steem:vote:votethisround}-\ref{alg:steem:vote:voterounds:end})
  that an honest player attempts to vote for as many posts as possible and
  spreads her votes with the maximum distance between them. The purpose of this
  is to efficiently utilize the available Voting Power to ``make her voice
  heard''. Also, efficiently using Voting Power on the Steemit website increases
  the voter's curation reward~\cite{curationRewards}.
\end{remark}

    \begin{theorem}
  \label{theorem:convergence:steem} \
  \begin{enumerate}
    \item \label{theorem:case:varstpow} If $\exists i \neq j \in
    \left[\playerlen\right]: \stpow_i \neq \stpow_j$ (i.e. if not all players
    have the same Steem Power) then Steemit does not $\left(\playerlen, \rounds,
    \postlen, 1\right)$-converge.
    \item If $\forall i \neq j \in \left[\playerlen\right], \stpow_i = \stpow_j$
    (i.e. if all players have the same Steem Power) and
    \begin{enumerate}
      \item \label{theorem:case:manyrounds} $\rounds - 1 \geq \left(\postlen -
      1\right) \ceil*{\frac{a+b}{\regen}}$ then Steemit $\left(\playerlen,
      \rounds, \postlen, \postlen\right)$-converges.
      \item \label{theorem:case:fewrounds} $\rounds - 1 < \left(\postlen -
      1\right) \ceil*{\frac{a+b}{regen}}$ then Steemit does not
      $\left(\playerlen, \rounds,\postlen, 1\right)$-converge.
    \end{enumerate}
  \end{enumerate}
\end{theorem}

    \begin{proofsketch}
  When $\stpowvec$ is not constant, we build a post list where the most liked
  post is not preferred by rich players and thus is not placed at the top. For a
  constant $\stpowvec$, when $\rounds - 1 \geq \left(\postlen -
  1\right)\ceil*{\frac{a +b}{\regen}}$, there are enough rounds to ensure full
  regeneration of every player's Voting Power between two votes and thus the
  resulting post list reflects the true preferences of the players. In the
  opposite case, we can always craft a post list that exploits the fact that
  some votes are cast with reduced Voting Power in order to trick the system
  into placing a wrong post in the top position.
\end{proofsketch}

    See Appendix~\ref{appendix:proof} for full proof.

    \begin{corollary}
  \label{corollary:convergence:steem}
  The Steemit system parametrised according to Remark~\ref{remark:steemitparms},
  for any number of players $\playerlen \geq 2$, constant $\stpowvec$ and
  $\postlen \leq 70$ posts $\left(\playerlen, \rounds, \postlen,
  \postlen\right)$-converges. If $\postlen > 70$ or $\stpowvec$ is not constant,
  then there exists a list of posts such that the system does not
  $\left(\playerlen, \rounds, \postlen, 1\right)$-converge.
\end{corollary}

\section{Simulation}
    The previous outcomes are here complemented with experiments that verify our
    findings. We have implemented a simulation framework that realizes the
    execution of Steemit's post-voting system as defined above.

    In particular, we consider two separate scenarios: First, we simulate the
    case when all players follow the prescribed honest strategy of Steemit,
    investigating how the curation quality of the system varies with the number
    of voting rounds. We successfully reproduce the result of
    Theorem~\ref{theorem:convergence:steem}, which implies that the system
    converges perfectly when a sufficient number of voting rounds is permitted,
    but otherwise the resulting list of posts may have a 0-ideal rank, i.e. the
    top post may not have the best ideal score. Moreover, we compare our
    $t$-convergence metric with previously used metrics of convergence based on
    correlation demonstrating that they are very closely aligned.

    The second case measures how resilient is the curation quality of Steemit
    against dishonest agents. Since a creator is financially rewarded when her
    content is upvoted, she has incentive to promote her own posts. A
    combination of in-band methods (apart from striving to produce posts of
    higher quality) can help her to that end. Voting for one's own posts,
    refraining from voting posts created by others and obtaining
    Sybil~\cite{sybilattack} accounts that only vote for her posts are only an
    indicative subset. We thus examine the quality of the resulting list when
    certain users do not follow the honest protocol, but apply the
    aforementioned self-promoting methods. We observe that there exists a cutoff
    point above which a small increase in the number of selfish players has a
    detrimental effect to the $t$-ideal rank of the post voting system.
    Furthermore, we measure the number of positions on the list that the selfish
    post gains with respect to the number of selfish players.

  \subsection{Methodology}
    We leverage three metrics to compare the curated list with the ideal list:
    Kendall's Tau~\cite{kendall1955rank}, Spearman's
    Rho~\cite{spearman1904proof}, and $t$-ideal rank.

%
    In addition to the $t$-ideal rank and the rank correlation coefficients used
    in the first scenario, in the case of dishonest participants we include a
    metric that measures the gains of the selfish players. In particular, the
    metric is defined as the difference between the real position of the
    ``selfish'' post after the execution and its ranking according to the ideal
    order. We are thus able to measure how advantageous is for users to behave
    selfishly. Furthermore, $t$-ideal rank informs us how this behavior affects
    the overall quality of curation of the platform.

  \subsection{Execution}
    In all simulations, the likabilities of all ``honest'' posts have been drawn
    from the $\left[0, 1\right]$-uniform distribution and all players have Steem
    Power equal to 1; we leave the case of variable Steem Power as future work.

    \subsubsection*{Scenario A}
\begin{figure}[!htbp]
        \includegraphics[width=0.9\textwidth]{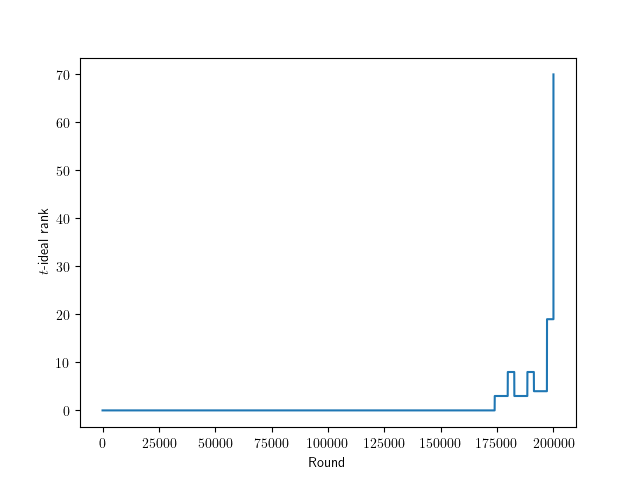}
        \caption{$t$-ideal rank evolution with 270 honest players, 70 posts and
        200.000 rounds}
        \label{fig:honest:70:tideal}
      \end{figure}

\begin{figure}[!htbp]
        \includegraphics[width=0.9\textwidth]{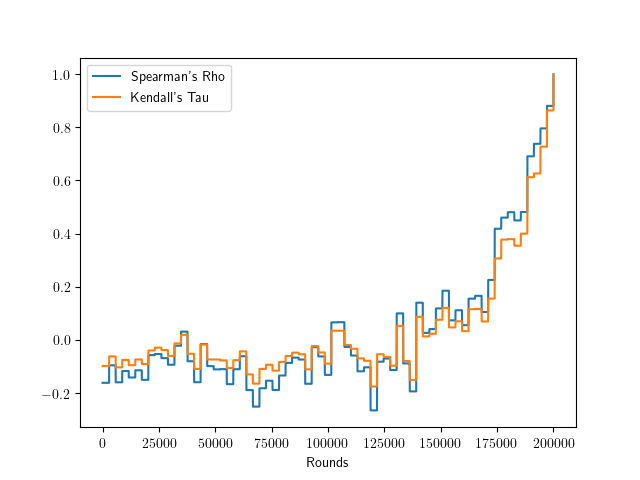}
        \caption{Kendall's Tau and Spearman's Rho evolution with 270 honest
        players, 70 posts and 200.000 rounds}
        \label{fig:honest:70:coeffs}
      \end{figure}

      As already mentioned, the results closely follow
      Theorem~\ref{theorem:convergence:steem}.
      Figures~\ref{fig:honest:70:tideal} and~\ref{fig:honest:70:coeffs} show the
      $t$-ideal rank and Kendall's Tau coefficient respectively when the number
      of rounds is enough for all votes to be cast with full Voting Power. In
      particular, the parameters used are $a = \frac{1}{50}, b = 10^{-4}, \regen
      = \frac{3}{5 \cdot 24 \cdot 60 \cdot 60}, \rounds = 200000, \attspan = 10,
      \playerlen = 270$ and $\postlen = 70$. (Observe that $\rounds - 1 >
      \left(\postlen - 1\right)\ceil*{\frac{a + b}{\regen}}$.)

      As we can see, all three measures show that the real list converges
      rapidly to the ideal order at the very end of the execution; meanwhile,
      the quality of the list improves very slowly.

\begin{figure}[!htbp]
        \includegraphics[width=0.9\textwidth]{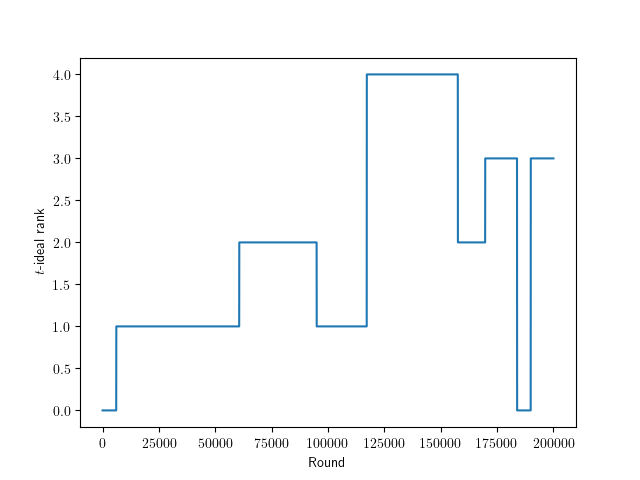}
        \caption{$t$-ideal rank evolution with 300 honest players, 100 posts and
        200.000 rounds}
        \label{fig:honest:100:tideal}
      \end{figure}

\begin{figure}[!htbp]
        \includegraphics[width=0.9\textwidth]{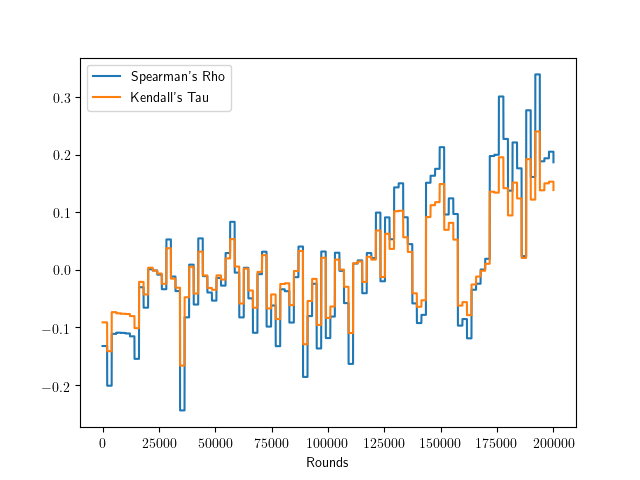}
        \caption{Kendall's Tau and Spearman's Rho evolution with 300 honest
        players, 100 posts and 200.000 rounds}
        \label{fig:honest:100:coeffs}
      \end{figure}

      Figures~\ref{fig:honest:100:tideal} and~\ref{fig:honest:100:coeffs} depict
      what happens when the rounds are not sufficient for all votes to be cast
      with full Voting Power. In particular, the corresponding simulation was
      executed with the same parameters, except for $\postlen = 100$ and
      $\playerlen = 300$. (Observe that $\rounds - 1 < \left(\postlen -
      1\right)\ceil*{\frac{a + b}{\regen}}$.)

      Here we see that at the end of the execution, only the first three posts
      are correctly ordered. Regarding the rest of the list, both Kendall's Tau
      and Spearman's Rho coefficients show that the order of the posts improves
      only slightly throughout the execution of the simulation.

    \subsubsection{Scenario B: Selfish users.}
\begin{figure}[!htbp]
        \includegraphics[width=0.9\textwidth]{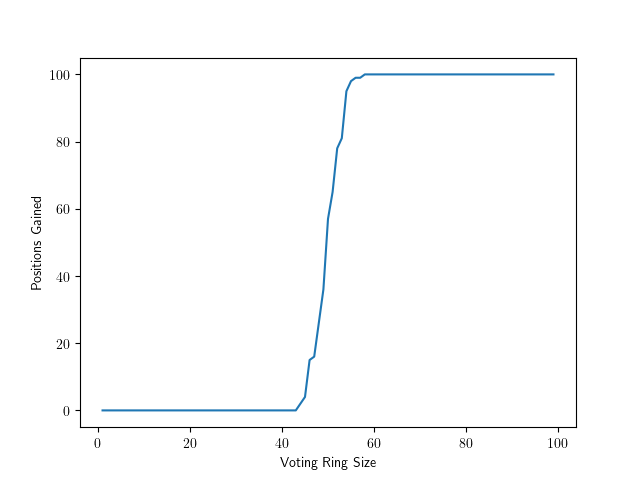}
        \caption{Positions gained by selfish post with 100 honest players, 100
        posts and 1 to 100 selfish players}
        \label{fig:selfish:gain}
      \end{figure}

\begin{figure}[!htbp]
        \includegraphics[width=0.9\textwidth]{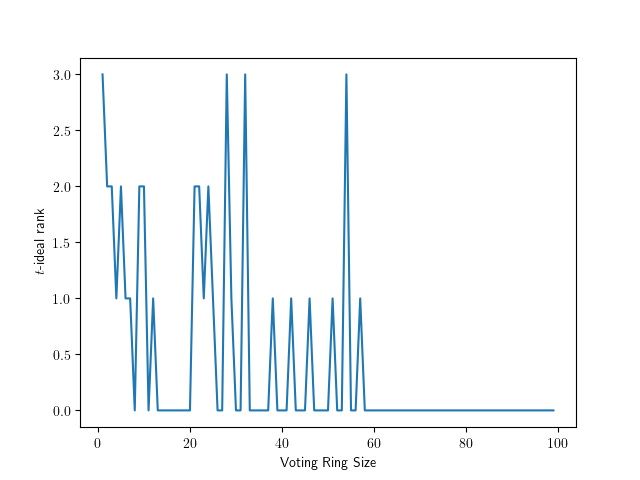}
        \caption{$t$-ideal rank with 100 honest players, 100 posts and 1 to 100
        selfish players}
        \label{fig:selfish:tideal}
      \end{figure}

      In order to understand how the presence of voting rings/Sybil accounts
      affects the curation quality, we simulate the execution of the game for
      various ring sizes, where ring members vote only for a particular,
      ``selfish'' post. We fix the rest of the system parameters to handicap the
      selfish post. In particular, the voting rounds are sufficient for all
      votes to be cast with full Voting Power, the likability of the selfish
      post is 0 for all players and it is initially placed at the bottom of the
      post list. Define the gain of the post of the selfish players as its ideal
      position minus its final position. Figure~\ref{fig:selfish:gain} shows the
      gain of the selfish post for a varying number of selfish players, from 1
      to 100.  Figure~\ref{fig:selfish:tideal} depicts the $t$-ideal rank of the
      resulting list at the same executions. The system parameters are
      $\playerlen = 101 .. 200, a = \frac{1}{50}, b = 10^{-4}, \regen =
      \frac{3}{5 \cdot 24 \cdot 60}, \attspan = 10, \rounds = 5000$.

\section{Summary and Future Work}

  We have defined an abstract post-voting system, along with a particularization
  inspired by the Steemit platform. We proved the exact conditions on the
  Steemit system parameters under which it successfully curates arbitrary lists
  of posts. We provided the results of simulations of the execution of the
  voting procedure under various conditions. Both cases with only honest and
  mixed honest and selfish players were simulated. We conclude that the Voting
  Power mechanism of Steem and the fact that self-voting is a profitable
  strategy may hurt curation quality.


  We have studied the curation properties of decentralized content curation
  platforms such as Steemit, obtaining new insights on the resilience of these
  systems. Some assumptions have been made in the presented model. Various
  relaxations of these assumptions constitute fertile ground for future work.
  First of all, the selfish strategy can be extended and refined in various
  ways. For example, voting rings can be allowed to create more than one posts
  in order to increase their rewards. Optimizing the number of posts and the
  vote allocation in this case would contribute towards a robust attack against
  the Steemit platform.

  Selfish behavior is considered only in the simulation. Our analysis can be
  augmented with a review of games with selfish players and voting rings.

  The addition of the economic factor invites the definition of utility
  functions and strategic behavior for the players. Its inclusion would imply
  the need for an expansion of our theorems and definitions to the strategic
  case, along with a full game-theoretic analysis. Furthermore, several possible
  refinements could be introduced; for example, the process of creating Sybil
  accounts could be associated with a monetary cost.

  Last but not least, in our model, posts are created only at the beginning of
  the execution. A dynamic model in which posts can be created at any time and
  the execution continues indefinitely (as is the case in a real-world UGC
  system) is also interesting as a future direction.

\begin{subappendices}
  \renewcommand{\thesection}{\Alph{section}}
  \section{Proof of Theorem~\ref{theorem:convergence:steem}: Steem convergence}
  \label{appendix:proof}
  \begin{proof}
  \begin{itemize}
    \item Statement~\ref{theorem:case:varstpow}: Reorder the players such that
    $\stpow_1 \geq \stpow_2 \geq \dots \geq \stpow_{\playerlen}$. Let $k =
    \min\limits_{j \in \left[\playerlen - 1\right]}\left\{\stpow_j \neq
    \stpow_{j+1}\right\}$. We first cover the case when $\attspan \geq 2$.

    Let\footnote{We thank Heng Guo from the University of Edinburgh for this
    counterexample.}
    \begin{gather*}
      \weakpost = (\underbrace{0, \dots, 0}_{k - 1}, 1, \underbrace{0,
      \dots, 0}_{\playerlen - k}) \\
      \strongpost = (\underbrace{0, \dots, 0}_{k - 1}, \frac{\stpow_k -
      \stpow_{k + 1}}{2\stpow_k}, 1, \underbrace{0, \dots, 0}_{\playerlen - k -
      1}) \\
      \nullpost = (\underbrace{0, \dots, 0}_{\playerlen}) \\
      \postlist = [\weakpost, \strongpost, \underbrace{\nullpost, \dots,
      \nullpost}_{\postlen - 2}] \enspace.
    \end{gather*}

    We first note that $\stpow_k > \stpow_{k+1} \geq 0 \Rightarrow 0 \leq
    \frac{\stpow_k - \stpow_{k + 1}}{2\stpow_k} \leq 1$, thus $\strongpost$ is a
    valid post. We then observe that
    \begin{gather*}
      \forall i \in \left\{3, \dots, \postlen\right\},
      \idsc{\postlist\left[i\right]} = 0 < \\
      <\idsc{\postlist\left[1\right]} = 1 < 1 + \frac{\stpow_k -
      \stpow_{k+1}}{2\stpow_k} = \idsc{\postlist\left[2\right]} \enspace,
    \end{gather*}
    thus $\forall \postlist'$ that contain the same posts as $\postlist$ and
    $\textsc{Ideal}^1\left(\postlist'\right)$ holds, it is
    $\postlist'\left[1\right] = \postlist\left[2\right]$.

    Since $\attspan \geq 2$, all players apart from $\player_{k + 1}$ vote for
    $\postlist\left[1\right]$ in the first round and for
    $\postlist\left[2\right]$ in the second, whereas $\player_{k + 1}$ votes for
    $\postlist\left[2\right]$ in the first round and for
    $\postlist\left[1\right]$ in the second. Thus the two first posts will have
    been voted by all players by the end of the second round and their score
    will not change until the execution completes. We have:

    \begin{gather*}
      \mathrm{sc}_2\left(\postlist\left[1\right]\right) =
      \mathrm{sc}_\rounds\left(\postlist\left[1\right]\right) = \\
      \sum\limits_{j = 1}^{k-1}\stpow_j b + \stpow_k \left(a + b\right) +
      \stpow_{k+1} \min{\left\{b, \votpowvecreg_{k+1,r_2}\right\}} +
      \sum\limits_{j = k + 2}^{\postlen}\stpow_j b \text{ and} \\
      \mathrm{sc}_2\left(\postlist\left[2\right]\right) =
      \mathrm{sc}_\rounds\left(\postlist\left[2\right]\right) = \\
      \sum\limits_{j = 1}^{k-1}\stpow_j \min{\{b, \votpowvecreg_{j,r_2}\}} + \\
      \stpow_k \min{\{a \frac{\stpow_k - \stpow_{k+1}}{2\stpow_k}
      \votpowvecreg_{k, r_2} + b, \votpowvecreg_{k, r_2}\}} + \stpow_{k+1}
      \left(a + b\right) + \\
      \sum\limits_{j = k + 2}^{\postlen}\stpow_j \min{\{b, \votpowvecreg_{j,
      r_2}\}} \Rightarrow
    \end{gather*}
    \begin{gather*}
      \mathrm{sc}_\rounds(\postlist\left[2\right]) \leq \\
      \sum\limits_{j = 1}^{k-1}\stpow_j b + \stpow_k (a \frac{\stpow_k -
      \stpow_{k+1}}{2\stpow_k} + b) + \stpow_{k+1} \left(a + b\right) +
      \sum\limits_{j = k + 2}^{\postlen}\stpow_j b \enspace.
    \end{gather*}

    In the case that $\votpowvecreg_{k+1, r_2} \geq b$, it is
    \begin{gather*}
      \mathrm{sc}_\rounds\left(\postlist\left[1\right]\right) = \sum\limits_{j =
      1}^{k-1}\stpow_j b + \stpow_k \left(a + b\right) + \stpow_{k+1} b +
      \sum\limits_{j = k + 2}^{\postlen}\stpow_j b > \\
      \sum\limits_{j = 1}^{k-1}\stpow_j b + \stpow_k (a \frac{\stpow_k -
      \stpow_{k+1}}{2\stpow_k} + b) + \stpow_{k+1} \left(a + b\right) +
      \sum\limits_{j = k + 2}^{\postlen}\stpow_j b \geq \\
      \mathrm{sc}_\rounds\left(\postlist\left[2\right]\right) \Rightarrow
      \mathrm{sc}_\rounds\left(\postlist\left[1\right]\right) >
      \mathrm{sc}_\rounds\left(\postlist\left[2\right]\right) \enspace,
    \end{gather*}
    thus $\textsc{Ideal}^1\left(\postlist'\right)$ does not hold.

    Since $\player_{k+1}$ does not vote in any round between $\round_1$ and
    $\round_2$, and $\round_2 \geq 2$, it is $\votpowvecreg_{k+1, r_2} \geq 1 -
    a - b + \regen$. Thus the case when $\votpowvecreg_{k+1, r_2} < b$ can
    happen only when $b > 1 - a - b + \regen \Leftrightarrow b > \frac{1 - a +
    \regen}{2}$. We now provide a counterexample for the case when $b > \frac{1
    - a + \regen}{2}$.

    Once more we order the players in descending Steem Power, like in the
    previous case. Once again $k = \min\limits_{j \in \left[\playerlen -
    1\right]}\left\{\stpow_j \neq \stpow_{j+1}\right\}$ and we only care for the
    case when $\attspan \geq 2$. Let $0 < \gamma < 1$ and
    \begin{gather*}
      \weakpost = (\underbrace{0, \dots, 0}_{k - 1}, 1, \frac{\gamma}{2},
      \underbrace{0, \dots, 0}_{\playerlen - k - 1}) \\
      \strongpost = (\underbrace{0, \dots, 0}_{k - 1}, \gamma, 1, \underbrace{0,
      \dots, 0}_{\playerlen - k - 1}) \\
      \nullpost = (\underbrace{0, \dots, 0}_{\playerlen}) \\
      \postlist = [\weakpost, \strongpost, \underbrace{\nullpost, \dots,
      \nullpost}_{\postlen - 2}] \enspace.
    \end{gather*}

    We observe that $\forall i \in \left\{3, \dots, \postlen\right\},
    \idsc{\postlist\left[i\right]} = 0 < \idsc{\postlist\left[1\right]} = 1 +
    \frac{\gamma}{2} < 1 + \gamma = \idsc{\postlist\left[2\right]}$, thus
    $\forall \postlist'$ that contain the same posts as $\postlist$ and
    $\textsc{Ideal}^1\left(\postlist'\right)$ holds, it is
    $\postlist'\left[1\right] = \postlist\left[2\right]$.

    Since $\attspan \geq 2$, all players apart from $\player_{k + 1}$ vote for
    $\postlist\left[1\right]$ in the first round and for
    $\postlist\left[2\right]$ in the second, whereas $\player_{k + 1}$ votes for
    $\postlist\left[2\right]$ in the first round and for
    $\postlist\left[1\right]$ in the second. Thus the two first posts will have
    been voted by all players by the end of the second round and their score
    will not change until the execution completes. We have:

    \begin{gather*}
      \mathrm{sc}_2\left(\postlist\left[1\right]\right) =
      \mathrm{sc}_\rounds\left(\postlist\left[1\right]\right) = \\
      \sum\limits_{j = 1}^{k-1}\stpow_j b + \stpow_k \left(a + b\right) +
      \stpow_{k+1} \votpowvecreg_{k+1,r_2} + \sum\limits_{j = k +
      2}^{\postlen}\stpow_j b \text{ and} \\
      \mathrm{sc}_2\left(\postlist\left[2\right]\right) =
      \mathrm{sc}_\rounds\left(\postlist\left[2\right]\right) = \\
      \sum\limits_{j = 1}^{k-1}\stpow_j \min{\{b, \votpowvecreg_{j, r_2}\}} +
      \stpow_k \votpowvecreg_{k, r_2} + \stpow_{k+1} \left(a + b\right) + \\
      \sum\limits_{j = k + 2}^{\postlen}\stpow_j \min{\{b, \votpowvecreg_{j,
      r_2}\}} \leq \\
      \sum\limits_{j = 1}^{k-1}\stpow_j b + \stpow_k \votpowvecreg_{k, r_2} +
      \stpow_{k+1} \left(a + b\right) +
      \sum\limits_{j = k + 2}^{\postlen}\stpow_j b \enspace.
    \end{gather*}

    We note that $\votpowvecreg_{k, r_2} = \votpowvecreg_{k+1, r_2}$ because
    both $\player_k$ and $\player_{k+1}$ vote with full Voting Power in the
    first round. Let $\mathrm{VP} = \votpowvecreg_{k, r_2}$. We have
    \begin{gather*}
      \stpow_k \left(a + b\right) + \stpow_{k+1} \mathrm{VP} > \stpow_k
      \mathrm{VP} + \stpow_{k+1} \left(a + b\right) \Leftrightarrow \\
      \stpow_k \left(a + b\right) + \stpow_{k+1} \mathrm{VP} - \stpow_k
      \mathrm{VP} - \stpow_{k+1} \left(a + b\right) > 0 \Leftrightarrow \\
      \left(a + b\right) \left(\stpow_k - \stpow_{k+1}\right) - \mathrm{VP}
      \left(\stpow_k - \stpow_{k+1}\right) > 0 \Leftrightarrow \\
      \left(\stpow_k - \stpow_{k+1}\right) \left(a + b - \mathrm{VP}\right) > 0
    \end{gather*}
    The last expression is true because $\stpow_k > \stpow_{k+1}$ and
    $\mathrm{VP} < b$, thus the first expression is true as well. We can then
    deduce that $\mathrm{sc}_\rounds\left(\postlist\left[1\right]\right) >
    \mathrm{sc}_\rounds\left(\postlist\left[2\right]\right)$, thus
    $\textsc{Ideal}^1\left(\postlist'\right)$ does not hold.

    We conclude the treatment of Statement~\ref{theorem:case:varstpow} by
    covering the case when $\attspan = 1$. Consider the following list of posts:
    \begin{gather*}
      \weakpost = \left(\frac{\stpow_1 - \stpow_\playerlen}{\stpow_1 -
      \stpow_\playerlen}, \dots, \frac{\stpow_i - \stpow_\playerlen}{\stpow_1 -
      \stpow_\playerlen}, \dots, \frac{\stpow_\playerlen -
      \stpow_\playerlen}{\stpow_1 - \stpow_\playerlen}\right) \\
      \strongpost = \\
      \left(\frac{\stpow_\playerlen - \stpow_\playerlen}{\stpow_1 -
      \stpow_\playerlen}, \dots, \frac{\stpow_{\playerlen - i + 1} -
      \stpow_\playerlen}{\stpow_1 - \stpow_\playerlen}, \dots, \frac{\stpow_{k +
      2} - \stpow_\playerlen}{\stpow_1 - \stpow_\playerlen}, \right. \\
      \left. \min{\left\{1, \frac{\stpow_{k + 1} - \stpow_\playerlen}{\stpow_1 -
      \stpow_\playerlen} + \frac{\left(\stpow_k -
      \stpow_{k+1}\right)^2}{2\stpow_{\playerlen - k}\max{\left\{1, \stpow_1 -
      \stpow_\playerlen\right\}}}\right\}},\right. \\
      \left.\frac{\stpow_{k} - \stpow_\playerlen}{\stpow_1 - \stpow_\playerlen},
      \dots, \frac{\stpow_1 - \stpow_\playerlen}{\stpow_1 -
      \stpow_\playerlen}\right) \\
      \nullpost = \left(\underbrace{0, \dots, 0}_{\playerlen}\right) \\
      \postlist = \left[\weakpost, \strongpost, \underbrace{\nullpost, \dots,
      \nullpost}_{\postlen - 2}\right] \enspace.
    \end{gather*}
    All the aforementioned posts are valid, since $\forall i \in
    \left[\playerlen\right], \stpow_1 \geq \stpow_i \geq \stpow_\playerlen
    \Rightarrow 1 \geq \frac{\stpow_i - \stpow_\playerlen}{\stpow_1 -
    \stpow_\playerlen} \geq 0$. Also $\frac{\left(\stpow_k -
    \stpow_{k+1}\right)^2}{2\stpow_{\playerlen - k}\max{\left\{1, \stpow_1 -
    \stpow_\playerlen\right\}}} > 0$, which ensures that
    $\strongpost_{\playerlen - k} > 0$. We also observe that $\stpow_{k+1} <
    \stpow_k \leq \stpow_1 \Rightarrow \frac{\stpow_{k+1} -
    \stpow_\playerlen}{\stpow_1 - \stpow_\playerlen} < 1$, thus
    $\strongpost_{\playerlen - k} > \frac{\stpow_{k+1} -
    \stpow_\playerlen}{\stpow_1 - \stpow_\playerlen}$.

    Regarding the ideal scores, we have
    \begin{gather*}
      \forall i \in \left\{3, \dots, \postlen\right\},
      \idsc{\postlist\left[i\right]} = 0 \enspace, \\
      \idsc{\postlist\left[1\right]} = \sum\limits_{i = 1}^\playerlen
      \frac{\stpow_i - \stpow_\playerlen}{\stpow_1 - \stpow_\playerlen}
      \enspace, \\
      \idsc{\postlist\left[2\right]} > \sum\limits_{i = 1}^\playerlen
      \frac{\stpow_{\playerlen - i + 1} - \stpow_\playerlen}{\stpow_1 -
      \stpow_\playerlen} = \sum\limits_{i = 1}^\playerlen
      \frac{\stpow_i - \stpow_\playerlen}{\stpow_1 - \stpow_\playerlen}
      \enspace,
    \end{gather*}
    Thus $\postlist\left[2\right]$ has the highest ideal score and $\forall
    \postlist'$ that contain the same posts as $\postlist$ and
    $\textsc{Ideal}^1\left(\postlist'\right)$ holds, it is
    $\postlist'\left[1\right] = \postlist\left[2\right]$.

    Since $\attspan = 1$, all players vote for all posts in the order they
    appear in the list of posts, thus $\postlist\left[1\right]$ is voted first,
    with full Voting Power by all players. $\postlist\left[2\right]$ will be
    voted next by all players with at most full Voting Power, thus
    $\mathrm{sc}_1\left(\postlist\left[1\right]\right) =
    \mathrm{sc}_\rounds\left(\postlist\left[1\right]\right)$ and
    $\mathrm{sc}_2\left(\postlist\left[2\right]\right) =
    \mathrm{sc}_\rounds\left(\postlist\left[2\right]\right)$. We will prove that
    $\mathrm{sc}_\rounds\left(\postlist\left[2\right]\right) <
    \mathrm{sc}_\rounds\left(\postlist\left[1\right]\right)$. It is
    \begin{gather*}
      \mathrm{sc}_\rounds\left(\postlist\left[1\right]\right) = \sum\limits_{i =
      1}^\playerlen \stpow_i\left(a \postlist\left[1\right]_i + b\right) =
      \sum\limits_{i = 1}^\playerlen \stpow_i\left(a \frac{\stpow_i -
      \stpow_\playerlen}{\stpow_1 - \stpow_\playerlen} + b\right) \enspace, \\
      \mathrm{sc}_\rounds\left(\postlist\left[2\right]\right) \leq
      \sum\limits_{i = 1}^\playerlen \stpow_i\left(a \postlist\left[2\right]_i +
      b\right) = \\
      \sum\limits_{\overset{i = 1}{i \neq \playerlen - k}}^\playerlen
      \stpow_i\left(a \frac{\stpow_{\playerlen - i + 1} -
      \stpow_\playerlen}{\stpow_1 - \stpow_\playerlen} + b\right) + \\
      \stpow_{\playerlen - k}\left(a \min\left\{1, \frac{\stpow_{k + 1} -
      \stpow_\playerlen}{\stpow_1 - \stpow_\playerlen} + \right.\right. \\
      \left.\left. \frac{\left(\stpow_k -
      \stpow_{k+1}\right)^2}{2\stpow_{\playerlen - k}\max{\left\{1, \stpow_1 -
      \stpow_\playerlen\right\}}}\right\} + b\right) \leq \\
      \sum\limits_{i = 1}^\playerlen \stpow_i\left(a \frac{\stpow_{\playerlen -
      i + 1} - \stpow_\playerlen}{\stpow_1 - \stpow_\playerlen} + b\right) + \\
      \stpow_{\playerlen - k} a \frac{\left(\stpow_k -
      \stpow_{k+1}\right)^2}{2\stpow_{\playerlen - k}\max{\left\{1, \stpow_1 -
      \stpow_\playerlen\right\}}} = A \enspace.
    \end{gather*}

    Since $\mathrm{sc}_\rounds\left(\postlist\left[2\right]\right) \leq A$, it
    is sufficient to prove $A <
    \mathrm{sc}_\rounds\left(\postlist\left[1\right]\right)$.
    \begin{gather*}
      A < \mathrm{sc}_\rounds\left(\postlist\left[1\right]\right)
      \Leftrightarrow \stpow_{\playerlen - k} a \frac{\left(\stpow_k -
      \stpow_{k+1}\right)^2}{2\stpow_{\playerlen - k}\max{\left\{1, \stpow_1 -
      \stpow_\playerlen\right\}}} + \\
      \sum\limits_{i = 1}^\playerlen \stpow_i\left(a \frac{\stpow_{\playerlen -
      i + 1} - \stpow_\playerlen}{\stpow_1 - \stpow_\playerlen} + b\right) - \\
      \sum\limits_{i = 1}^\playerlen \stpow_i\left(a \frac{\stpow_i -
      \stpow_\playerlen}{\stpow_1 - \stpow_\playerlen} + b\right) < 0
      \Leftrightarrow \\
      \frac{\left(\stpow_k - \stpow_{k+1}\right)^2}{2\max{\left\{1, \stpow_1 -
      \stpow_\playerlen\right\}}} + \\
      \sum\limits_{i = 1}^\playerlen \stpow_i \left(\frac{\stpow_{\playerlen - i
      + 1} - \stpow_\playerlen}{\stpow_1 - \stpow_\playerlen} - \frac{\stpow_i -
      \stpow_\playerlen}{\stpow_1 - \stpow_\playerlen}\right) < 0
      \Leftrightarrow \\
      \frac{\left(\stpow_k - \stpow_{k+1}\right)^2}{2\max{\left\{1, \stpow_1 -
      \stpow_\playerlen\right\}}} + \sum\limits_{i = 1}^\playerlen \stpow_i
      \frac{\stpow_{\playerlen - i + 1} - \stpow_i}{\stpow_1 -
      \stpow_\playerlen} < 0
    \end{gather*}
    It is
    \begin{equation}
      \label{conv:proof:varstpow:incfrac}
      \frac{\left(\stpow_k - \stpow_{k+1}\right)^2}{2\max{\left\{1, \stpow_1 -
      \stpow_\playerlen\right\}}} \leq \frac{\left(\stpow_k -
      \stpow_{k+1}\right)^2}{2 \left(\stpow_1 - \stpow_\playerlen\right)}
      \enspace.
    \end{equation}
    Furthermore,
    \begin{equation}
      \label{conv:proof:varstpow:split}
      \begin{gathered}
        \sum\limits_{i = 1}^\playerlen \stpow_i \left(\stpow_{\playerlen - i +
        1} - \stpow_i\right) = \\
        \sum\limits_{i = 1}^{\floor*{\frac{\playerlen}{2}}}
        \left(\stpow_i\left(\stpow_{\playerlen - i + 1} - \stpow_i\right) +
        \stpow_{\playerlen - i + 1}\left(\stpow_i - \stpow_{\playerlen - i +
        1}\right)\right) \enspace.
      \end{gathered}
    \end{equation}
    From~(\ref{conv:proof:varstpow:incfrac})
    and~(\ref{conv:proof:varstpow:split}),
    it suffices to prove that
    \begin{gather*}
      \frac{\left(\stpow_k - \stpow_{k+1}\right)^2}{2 \left(\stpow_1 -
      \stpow_\playerlen\right)} + \\
      \sum\limits_{i = 1}^{\floor*{\frac{\playerlen}{2}}}
      \frac{\stpow_i\left(\stpow_{\playerlen - i + 1} - \stpow_i\right) +
      \stpow_{\playerlen - i + 1}\left(\stpow_i - \stpow_{\playerlen - i +
      1}\right)}{\stpow_1 - \stpow_\playerlen} < 0 \enspace.
    \end{gather*}
    \begin{gather*}
      \frac{\left(\stpow_k - \stpow_{k+1}\right)^2}{2 \left(\stpow_1 -
      \stpow_\playerlen\right)} + \\
      \sum\limits_{i = 1}^{\floor*{\frac{\playerlen}{2}}}
      \frac{\stpow_i\left(\stpow_{\playerlen - i + 1} - \stpow_i\right) +
      \stpow_{\playerlen - i + 1}\left(\stpow_i - \stpow_{\playerlen - i +
      1}\right)}{\stpow_1 - \stpow_\playerlen} < 0 \Leftrightarrow \\
      \frac{\left(\stpow_k - \stpow_{k+1}\right)^2}{2} + \\
      \sum\limits_{i = 1}^{\floor*{\frac{\playerlen}{2}}}
      \left(\stpow_i\left(\stpow_{\playerlen - i + 1} - \stpow_i\right) -
      \stpow_{\playerlen - i + 1}\left(\stpow_{\playerlen - i + 1} -
      \stpow_i\right)\right) < 0 \Leftrightarrow \\
      \frac{\left(\stpow_k - \stpow_{k+1}\right)^2}{2} - \sum\limits_{i =
      1}^{\floor*{\frac{\playerlen}{2}}} \left(\stpow_i - \stpow_{\playerlen - i
      + 1}\right)^2 < 0
    \end{gather*}
    If $k \leq \floor*{\frac{\playerlen}{2}}$, then
    \begin{gather*}
      \frac{\left(\stpow_k - \stpow_{k+1}\right)^2}{2} - \sum\limits_{i =
      1}^{\floor*{\frac{\playerlen}{2}}} \left(\stpow_i - \stpow_{\playerlen - i
      + 1}\right)^2 = \\
      \frac{\left(\stpow_k - \stpow_{k+1}\right)^2}{2} - \left(\stpow_k -
      \stpow_{\playerlen - k + 1}\right)^2 - \sum\limits_{\overset{i = 1}{i \neq
      k}}^{\floor*{\frac{\playerlen}{2}}} \left(\stpow_i - \stpow_{\playerlen -
      i + 1}\right)^2
    \end{gather*}
    It is
    \begin{gather*}
      - \sum\limits_{\overset{i = 1}{i \neq k}}^{\floor*{\frac{\playerlen}{2}}}
      \left(\stpow_i - \stpow_{\playerlen - i + 1}\right)^2 \leq 0 \text{ and}
      \\
      k \leq \floor*{\frac{\playerlen}{2}} \Rightarrow \playerlen - k + 1 \geq k
      + 1 \Rightarrow \stpow_{\playerlen - k + 1} \leq \stpow_{k + 1}
      \Rightarrow \\
      \left(\stpow_k - \stpow_{k + 1}\right)^2 \geq \left(\stpow_k -
      \stpow_{\playerlen - k + 1}\right)^2 \overset{\stpow_k >
      \stpow_{\playerlen - k + 1}}{\Rightarrow} \\
      \left(\stpow_k - \stpow_{k + 1}\right)^2 > \frac{\left(\stpow_k -
      \stpow_{\playerlen - k + 1}\right)^2}{2} \Rightarrow \\
      \frac{\left(\stpow_k - \stpow_{\playerlen - k + 1}\right)^2}{2} -
      \left(\stpow_k - \stpow_{k + 1}\right)^2 < 0
    \end{gather*}
    We deduce that
    \begin{equation*}
      \frac{\left(\stpow_k - \stpow_{k+1}\right)^2}{2} - \sum\limits_{i =
      1}^{\floor*{\frac{\playerlen}{2}}} \left(\stpow_i - \stpow_{\playerlen - i
      + 1}\right)^2 < 0
    \end{equation*}

    Else if $k > \floor*{\frac{\playerlen}{2}}$, then
    \begin{gather*}
      \frac{\left(\stpow_k - \stpow_{k+1}\right)^2}{2} - \sum\limits_{i =
      1}^{\floor*{\frac{\playerlen}{2}}} \left(\stpow_i - \stpow_{\playerlen - i
      + 1}\right)^2 = \\
      \frac{\left(\stpow_k - \stpow_{k+1}\right)^2}{2} -
      \left(\stpow_{\playerlen - k} - \stpow_{k + 1}\right)^2 -
      \sum\limits_{\overset{i = 1}{i \neq \playerlen -
      k}}^{\floor*{\frac{\playerlen}{2}}} \left(\stpow_i - \stpow_{\playerlen -
      i + 1}\right)^2
    \end{gather*}
    It is
    \begin{gather*}
      - \sum\limits_{\overset{i = 1}{i \neq \playerlen -
      k}}^{\floor*{\frac{\playerlen}{2}}} \left(\stpow_i - \stpow_{\playerlen -
      i + 1}\right)^2 \leq 0 \text{ and} \\
      k > \floor*{\frac{\playerlen}{2}} \Rightarrow \playerlen - k < k
      \Rightarrow \stpow_{\playerlen - k} \geq \stpow_k \Rightarrow \\
      \left(\stpow_{\playerlen - k} - \stpow_{k + 1}\right)^2 \geq
      \left(\stpow_k - \stpow_{k + 1}\right)^2 \overset{\stpow_k > \stpow_{k +
      1}}{\Rightarrow} \\
      \left(\stpow_{\playerlen - k} - \stpow_{k + 1}\right)^2 >
      \frac{\left(\stpow_k - \stpow_{k + 1}\right)^2}{2} \Rightarrow \\
      \frac{\left(\stpow_k - \stpow_{k + 1}\right)^2}{2} -
      \left(\stpow_{\playerlen - k} - \stpow_{k + 1}\right)^2 < 0
    \end{gather*}
    We deduce that
    \begin{equation*}
      \frac{\left(\stpow_k - \stpow_{k+1}\right)^2}{2} - \sum\limits_{i =
      1}^{\floor*{\frac{\playerlen}{2}}} \left(\stpow_i - \stpow_{\playerlen - i
      + 1}\right)^2 < 0
    \end{equation*}
    We have concluded that in every case
    $\mathrm{sc}_\rounds\left(\postlist\left[2\right]\right) <
    \mathrm{sc}_\rounds\left(\postlist\left[1\right]\right)$, thus
    $\textsc{Ideal}^1\left(\postlist'\right)$ does not hold.

    \item Statement~\ref{theorem:case:manyrounds}: Suppose that
    \begin{equation}
      \label{conv:proof:rev}
      \rounds - 1 \geq \left(\postlen - 1\right)\ceil*{\frac{a+b}{\regen}}
      \enspace.
    \end{equation}
    Observe that
    \begin{equation}
      \label{conv:proof:integer}
      \left(\ref{conv:proof:rev}\right) \Rightarrow \frac{\rounds - 1}{\postlen
      - 1} \geq \ceil*{\frac{a+b}{\regen}}
      \underset{\text{integer}}{\overset{\text{rhs}}{\Rightarrow}}
      \floor*{\frac{\rounds - 1}{\postlen - 1}} \geq \ceil*{\frac{a+b}{\regen}}
      \enspace.
    \end{equation}
    Let $\pid \in \left[\playerlen\right]$. From~(\ref{conv:proof:rev}) we
    deduce that $\rounds \geq \postlen$ and according to \textsc{VoteThisRound}
    in Algorithm~\ref{alg:steem:vote}, $u_{\pid}$ votes non-null in rounds
    $\left(\round_1, \dots, \round_\postlen\right)$ with $r_i = \floor*{\left(i
    - 1\right)\frac{\rounds - 1}{\postlen - 1}} + 1$. We define the following:
    \begin{gather*}
      k \in \mathbb{N}, w \in \mathbb{R} \enspace, \\
      n \in \mathbb{Z}, p \in \left[0, 1\right) : \left(k - 1\right)w = n + p
      \enspace, \\
      m \in \mathbb{Z}, q \in \left[0, 1\right) : w = m + q \enspace.
    \end{gather*}
    We have
    \begin{gather}
      \label{conv:proof:step:kw}
      \floor*{\left(k - 1\right)w} = n \enspace, \\
      \label{conv:proof:step:almostkw}
      \floor*{kw} =
      \begin{cases}
        n + m, & p + q < 1 \\
        n + m + 1, & p + q \geq 1 \text{ (impossible if } p = 0\text{)}
      \end{cases} \\
      \label{conv:proof:step:w:floor}
      \floor*{w} = m \\
      \label{conv:proof:step:w:ceil}
      \ceil*{w} =
      \begin{cases}
        m, & p = 0 \\
        m + 1, & p > 0
      \end{cases}
    \end{gather}
    \begin{equation}
      \label{conv:proof:step:floors}
      \begin{gathered}
        (\ref{conv:proof:step:kw}), (\ref{conv:proof:step:almostkw}),
        (\ref{conv:proof:step:w:floor}), (\ref{conv:proof:step:w:ceil}), p + q <
        2 \Rightarrow \\
        \floor*{kw} \in \left\{\floor*{\left(k - 1\right)w} + \floor*{w},
        \floor*{\left(k - 1\right)w} + \ceil*{w}\right\}
      \end{gathered}
    \end{equation}

    From~(\ref{conv:proof:step:floors}) we deduce that
    \begin{equation}
      \label{conv:proof:step}
      \forall i \in \left[\postlen\right] \setminus \left\{1\right\}, r_i
      \in \left\{r_{i - 1} + \floor*{\frac{\rounds - 1}{\postlen - 1}}, r_{i -
      1} + \ceil*{\frac{\rounds - 1}{\postlen - 1}}\right\} \enspace.
    \end{equation}
    From (\ref{conv:proof:integer}) and (\ref{conv:proof:step}) we have that
    $\forall i \in \left[\postlen - 1\right], r_{i+1} - r_i \geq
    \ceil*{\frac{a+b}{\regen}}$. We will now prove by induction that $\forall i
    \in \left[\postlen\right], \votpowvec_{\pid, r_i} = 1$.

    \begin{itemize}
      \item For $i = 1, \votpowvec_{\pid, 1} = 1$
      (Algorithm~\ref{alg:steem:init}, line~\ref{alg:steem:init:vp}).
      \item Let $\votpowvec_{\pid, r_i} = 1$. Until $r_{i + 1}$, a single
      non-null vote is cast by $u_\pid$, which reduces $\votpowvec_{\pid}$ by at
      most $a + b$ (Algorithm~\ref{alg:steem:handlevote},
      line~\ref{alg:steem:handlevote:cost:start}) and at least $\ceil*{\frac{a +
      b}{\regen}}$ regenerations, each of which replenishes $\votpowvec_{\pid}$
      by $\regen$. Thus
      \begin{equation*}
        \votpowvec_{\pid, r_{i + 1}} \geq \min{\left\{\votpowvec_{\pid, r_i} - a
        - b + \regen\ceil*{\frac{a + b}{\regen}}, 1\right\}} \geq 1 \enspace.
      \end{equation*}
      But $\votpowvec_{\pid}$ cannot exceed 1
      (line~\ref{alg:steem:handlevote:regen}), thus $\votpowvec_{\pid, r_{i +
      1}} = 1$.
    \end{itemize}
    Since the above holds for every $\pid \in \left[\playerlen\right]$, it holds
    that at the end of the execution, all votes have been cast with full Voting
    Power, thus $\forall i \in \left[\postlen\right],
    \mathrm{sc}_{\rounds}\left(\postlist\left[i\right]\right) =
    \playerlen b + a \sum\limits_{\pid = 1}^\playerlen
    \postlist\left[i\right]_{\pid}$ and the posts in $\postlist_R$ are sorted by
    decreasing score (Algorithm~\ref{alg:steem:handlevote},
    line~\ref{alg:steem:handlevote:order}). We observe that
    \begin{gather*}
      \forall i \neq j \in \left[\postlen\right], \idsc{\postlist\left[i\right]}
      > \idsc{\postlist\left[j\right]} \Rightarrow \\
      \sum\limits_{\pid = 1}^\playerlen \postlist\left[i\right]_\pid >
      \sum\limits_{\pid = 1}^\playerlen \postlist\left[j\right]_\pid \Rightarrow
      \\
      \playerlen b + a \sum\limits_{\pid = 1}^\playerlen
      \postlist\left[i\right]_\pid > \playerlen b + a \sum\limits_{\pid =
      1}^\playerlen \postlist\left[j\right]_\pid \enspace.
    \end{gather*}
    Thus all posts will be ordered according to their ideal scores; put
    otherwise, $\textsc{IdealScore}^M\left(\postlist_\rounds\right)$ holds.

    \item Statement~\ref{theorem:case:fewrounds}: Suppose that
    \begin{equation}
      \label{conv:proof:fewrounds}
      \rounds - 1 < \left(\postlen - 1\right)\ceil*{\frac{a+b}{\regen}}
      \enspace.
    \end{equation}

    Several lists of posts will be defined in the rest of the proof. Given that,
    when all players are honest, the creator of a post is irrelevant, we omit
    the creator from the definition of posts to facilitate the exposition. Thus
    every post will be defined as a tuple of likabilities.

    First, we consider the case when
    \begin{equation}
      \label{conv:proof:smallattspanrounds}
      \attspan + \rounds \leq \postlen \enspace.
    \end{equation}
    In this case, no player can ever vote for the last post, as we will show
    now. First of all, $(\ref{conv:proof:smallattspanrounds}) \Rightarrow
    \rounds < \postlen$, thus all players cast $\rounds$ votes in total. Let
    $\pid \in \playerlen, i \in \left[\rounds\right]$ and $v_{\pid, i}$ the
    index of the last post that has ever been in $\player_\pid$'s attention
    span until the end of round $i$, according to the ordering of $\postlist$.
    It is $v_{\pid, 1} = \attspan$ and $\forall i \in \left[\rounds\right]
    \setminus \left\{1\right\}, v_{\pid, i} = v_{\pid, i - 1} + 1$, since in
    every round $\player_\pid$ votes for a single post and the first unvoted
    post of the list is added to their attention span. Note that, since this
    mechanism is the same for all players, the same unvoted post is added to
    all players' attention span at every round. Thus $\forall \pid \in
    \playerlen, v_{\pid, \rounds} = \attspan + \rounds - 1
    \overset{\left(\ref{conv:proof:smallattspanrounds}\right)}{<} \postlen$.
    We deduce that no player has ever the chance to vote for the last post.

    The above observation naturally leads us to the following counterexample:
    Let
    \begin{gather*}
      \strongpost = \left(\underbrace{1, \dots, 1}_{\playerlen}\right) \\
      \nullpost = \left(\underbrace{0, \dots, 0}_{\playerlen}\right) \\
      \postlist = \left[\underbrace{\nullpost, \dots, \nullpost}_{\postlen -
      1}, \strongpost\right]
    \end{gather*}

    $\forall i \in \left[\postlen - 1\right],$ it is
    $\idsc{\postlist\left[\postlen\right]} >
    \idsc{\postlist\left[i\right]}$, thus $\forall \postlist'$ that
    contain the same posts as $\postlist$ and
    $\textsc{Ideal}^1\left(\postlist'\right)$ holds, it is
    $\postlist'\left[1\right] = \postlist\left[\postlen\right]$. However,
    since the last post is not voted by any player and the first post is voted
    by at least one player, it is
    $\mathrm{sc}_\rounds\left(\postlist\left[1\right]\right) >
    \mathrm{sc}_\rounds\left(\postlist\left[\postlen\right]\right)$, thus
    $\textsc{Ideal}^1\left(\postlist_\rounds\right)$ does not hold.

    We now move on to the case when $\attspan + \rounds > \postlen$. Let
    $\votenum = \min{\left\{\rounds, \postlen\right\}}$. Each player casts
    exactly $\votenum$ votes. Consider $\postlist^1 = 1^{\postlen \times
    \playerlen}$ and $\pid \in \left[\playerlen\right]$. Let
    \begin{equation*}
      i \in \left[\votenum\right] :\left(\votpowvecreg_{\pid, \round_i} < 1
      \wedge \nexists i' < i: \votpowvecreg_{\pid, \round_{i'}} < 1\right)
      \enspace,
    \end{equation*}
    i.e. $i$ is the first round in which $u_\pid$ votes with less than full
    Voting Power. Such a round exists in every case as we will show now. Note
    that, since the first round is a voting round and the Voting Power of all
    players is full at the beginning, if $i$ exists it is $i \geq 2$.

    \begin{itemize}
      \item If $\rounds \geq \postlen$, it is $\votenum = \postlen$. \\
      If $\nexists i \in \left[\postlen\right] :\left(\votpowvecreg_{\pid,
      \round_i} < 1 \wedge \nexists i' < i: \votpowvecreg_{\pid, \round_{i'}} <
      1\right)$, then $\forall i \in \left[\postlen\right], \votpowvecreg_{\pid,
      \round_i} = 1 \Rightarrow \forall i \in \left[\postlen\right] \setminus
      \left\{1\right\}, \round_i \geq \round_{i - 1} + \ceil*{\frac{a +
      b}{\regen}}$ to have enough rounds to replenish the Voting Power after a
      full-weight, full-Voting Power vote. Thus $\round_\postlen \geq 1 +
      \left(\postlen - 1\right)\ceil*{\frac{a + b}{\regen}} > \rounds$,
      contradiction.

      \item If $\rounds < \postlen$, every player votes on all rounds, thus $r_2
      = 2$. Note that
      \begin{equation}
        \label{conv:proof:normalregen}
        \ceil*{\frac{a + b}{\regen}} \geq 2 \Rightarrow \frac{a + b}{\regen} >
        1 \Rightarrow a + b > \regen\enspace.
      \end{equation}
      Thus $\forall \pid \in \left[\playerlen\right], \votpowvecreg_{\pid, r_2}
      = 1 - a - b + \regen \overset{(\ref{conv:proof:normalregen})}{<} 1$, thus
      $i = 2$.
    \end{itemize}
    We proved that $i$ exists. Since all players follow the same voting pattern,
    the Voting Power of all players in each round is the same. Let $\rvp =
    \votpowvecreg_{1, \round_i}$. Assume that $\attspan < i \vee i > 2$. We
    cover the case where $\attspan \geq i \wedge i = 2$ later. In case
    $\playerlen$ is even, let $0 < \gamma < 0, 0 < \epsilon < \gamma\left(1 -
    \rvp\right)$,

    \begin{gather*}
      \weakpost = \left(\underbrace{1, \dots, 1}_{\playerlen/2},
      \underbrace{\gamma - \epsilon, \dots, \gamma -
      \epsilon}_{\playerlen/2}\right) \enspace, \\
      \strongpost = \left(\underbrace{\gamma, \dots, \gamma}_{\playerlen/2},
      \underbrace{1, \dots, 1}_{\playerlen/2}\right) \enspace, \\
      \nullpost = \left(\underbrace{0, \dots, 0}_{\playerlen}\right) \enspace,
      \\
      \postlist = \left[\underbrace{\weakpost, \dots, \weakpost}_{i - 1},
      \strongpost, \underbrace{\nullpost, \dots, \nullpost}_{\postlen -
      i}\right] \enspace.
    \end{gather*}

    First of all, it is
    \begin{gather*}
      \forall j \in \left[i - 1\right], \idsc{\postlist\left[j\right]} =
      \frac{\playerlen}{2}\left(1 + \gamma - \epsilon\right) < \\
      < \frac{\playerlen}{2}\left(1 + \gamma\right) =
      \idsc{\postlist\left[i\right]}
    \end{gather*}
    and $\forall j \in \left\{i + 1, \dots,
    \postlen\right\}, \idsc{\postlist\left[j\right]} = 0 <
    \idsc{\postlist\left[i\right]}$, thus the strong post has strictly the
    highest ideal score of all posts and as a result, $\forall \postlist'$ that
    contains the same posts as $\postlist$ and
    $\textsc{Ideal}^1\left(\postlist'\right)$ holds, it is
    $\postlist'\left[1\right] = \postlist\left[i\right]$.

    We observe that all players like both weak and strong posts more than null
    posts, thus no player will vote for a null post unless her attention span
    contains only null posts. This can happen in two cases: First, if the player
    has not yet voted for all non-null posts, but the first $\attspan$ posts of
    the list, excluding already voted posts, are null posts. Second, if the
    player has already voted for all non-null posts. For a null post to rank
    higher than a non-null one, it must be true that there exists one player
    that has cast the first vote for the null post. However, since the null
    posts are initially at the bottom of the list and it is impossible for a
    post to improve its ranking before it is voted, we deduce that this first
    vote can be cast only after the voter has voted for all non-null posts. We
    deduce that all players vote for all non-null posts before voting for any
    null post.

    We will now see that the first $\frac{\playerlen}{2}$ players vote first for
    all weak posts and then for the strong post. These players like the weak
    posts more than the strong post. As we saw, they will not vote any null
    post before voting for all non-null ones. If $\attspan > 1$ they vote for
    the strong post only when all other posts in their attention span are null
    ones and thus they will have voted for all weak posts already. If $\attspan
    = 1$ and since no post can increase its position before being voted, the
    strong post will become ``visible'' for all players only once they have
    voted for all weak posts. Thus in both cases the first
    $\frac{\playerlen}{2}$ players vote for the strong post only after they have
    voted for all weak posts first.

    The two previous results combined prove that the first
    $\frac{\playerlen}{2}$ players vote for the strong post in round $\round_i$
    exactly. We also observe that these players have experienced the exact same
    Voting Power reduction and regeneration as in the case of $\postlist^1$
    since they voted only for posts with likability 1, thus in round $\round_i$
    their Voting Power after regeneration is exactly the same as in the case of
    $\postlist^1: \forall \pid \in \left[\frac{\playerlen}{2}\right],
    \votpowvecreg_{\pid, \round_i} = \rvp$.

    We observe that the first $\frac{\playerlen}{2}$ players vote for all weak
    posts with full Voting Power. As for the last $\frac{\playerlen}{2}$
    players, we observe that, if $\attspan < i$, they all vote for the first
    weak post of the list in the first round, and thus with full Voting Power. If
    $\attspan \geq i$ and $i > 2$, they vote for the strong post in the first
    round and for the first weak post in $r_2$ with full Voting Power. Thus in
    all cases the last $\frac{\playerlen}{2}$ players vote for the first weak
    post with full Voting Power. Therefore, the score of the first weak post at
    the end of the execution is
    $\mathrm{sc}_{\rounds}\left(\postlist\left[1\right]\right) =
    \frac{\playerlen}{2}\left(a + b\right) +
    \frac{\playerlen}{2}\left(\left(\gamma - \epsilon\right) a + b\right)$.

    On the other hand, at the end of the execution the strong post has been
    voted by the first $\frac{\playerlen}{2}$ players with $\rvp$
    Voting Power and by the last $\frac{\playerlen}{2}$ players with at most
    full Voting Power, thus its final score will be at most
    $\mathrm{sc}_{\rounds}\left(\postlist\left[i\right]\right) \leq
    \frac{\playerlen}{2}\left(\rvp \cdot \gamma a + b\right) +
    \frac{\playerlen}{2}\left(a + b\right)$. It is
    \begin{gather*}
      \epsilon < \gamma\left(1 - \rvp\right) \Rightarrow \\
      \frac{\playerlen}{2}\left(\rvp \cdot \gamma a + b\right) +
      \frac{\playerlen}{2}\left(a + b\right) < \frac{\playerlen}{2}\left(a +
      b\right) + \frac{\playerlen}{2}\left(\left(\gamma - \epsilon\right) a +
      b\right) \Rightarrow \\
      \mathrm{sc}_{\rounds}\left(\postlist\left[i\right]\right) <
      \mathrm{sc}_{\rounds}\left(\postlist\left[1\right]\right) \enspace.
    \end{gather*}

    Thus $\postlist_{\rounds}\left[1\right] \neq \postlist\left[i\right]$ and
    $\mathrm{Ideal}^1\left(\postlist_{\rounds}\right)$ does not hold.

    As for the case when $\playerlen$ is odd, let $0 < \epsilon <
    \gamma\frac{\playerlen - 3}{\playerlen - 1}\left(1 - \rvp\right)$. In this
    case, we assume that the likability of the first $i$ posts (weak and strong)
    for the additional player is $\gamma$, whereas the likability of the last
    $\postlen - i$ posts (the null posts) is 0. This means that the additional
    player votes first for the weak and strong posts and then for the null
    posts. The rest of the likabilities remain as in the case when $\playerlen$
    is even. We observe that the ideal score of the strong post is still
    strictly higher than the rest. Furthermore, since the additional player
    votes for the first weak post within the first $i$ voting rounds, her Voting
    Power at the time of this vote will be at least $\rvp$. We thus have the
    following bounds for the scores:

    \begin{gather*}
      \mathrm{sc}_{\rounds}\left(\postlist\left[i\right]\right) \leq
      \frac{\playerlen - 1}{2}\left(\rvp \cdot \gamma a + b\right) +
      \frac{\playerlen - 1}{2}\left(a + b\right) + \gamma a + b \enspace, \\
      \mathrm{sc}_{\rounds}\left(\postlist\left[1\right]\right) \geq
      \frac{\playerlen - 1}{2}\left(a + b\right) + \frac{\playerlen -
      1}{2}\left(\left(\gamma - \epsilon\right) a + b\right) + \rvp \cdot \gamma
      a + b \enspace.
    \end{gather*}

    Given the bounds of $\epsilon$, it is
    $\mathrm{sc}_{\rounds}\left(\postlist\left[i\right]\right) <
    \mathrm{sc}_{\rounds}\left(\postlist\left[1\right]\right)$, thus
    $\mathrm{Ideal}^1\left(\postlist_{\rounds}\right)$ does not hold.

    We finally cover the previously untreated edge case where $\attspan \geq i
    \wedge i = 2$. $\rvp$ is defined like before. We first consider the case
    when $\playerlen$ is even and greater than 2: $\exists k \in \mathbb{N}
    \setminus \left\{0, 1\right\}: \playerlen = 2k$. Let $0 < \gamma < 1, 0 <
    \epsilon < 2 \gamma \frac{1 - \rvp}{\left(k - 1\right)\rvp}$,
    \begin{gather*}
      \weakpost = \left(\underbrace{1, \dots, 1}_{k - 1}, \underbrace{\gamma -
      \epsilon, \dots, \gamma - \epsilon}_{k - 1}, \gamma, \gamma\right)
      \enspace, \\
      \strongpost = \left(\underbrace{\gamma, \dots, \gamma}_{k - 1},
      \underbrace{1, \dots, 1}_{k - 1}, \gamma, \gamma\right) \enspace, \\
      \postlist = \left[\weakpost, \strongpost, \underbrace{\nullpost, \dots,
      \nullpost}_{\postlen - 2}\right] \enspace.
    \end{gather*}
    We first observe that
    \begin{gather*}
      \forall j \in \left\{3, \dots, \postlen\right\},
      \idsc{\postlist\left[j\right]} = 0 < \\
      < \idsc{\postlist\left[1\right]} = k - 1 + \left(k - 1\right)\left(\gamma
      - \epsilon\right) + 2\gamma = \\
      = k - 1 + \left(k + 1\right)\gamma - \left(k - 1\right)\epsilon < \\
      < k - 1 + \left(k + 1\right)\gamma = \idsc{\postlist\left[2\right]}
      \enspace,
    \end{gather*}
    thus the strong post has strictly the highest ideal score of all posts and
    as a result, $\forall \postlist'$ that contains the same posts as
    $\postlist$ and $\textsc{Ideal}^1\left(\postlist'\right)$ holds, it is
    $\postlist'\left[1\right] = \postlist\left[2\right]$.

    The first $k - 1$ and the last two players vote first for
    $\postlist\left[1\right]$ and then for $\postlist\left[2\right]$, whereas
    players $k, \dots, 2k - 2$ vote first for $\postlist\left[2\right]$ and then
    for $\postlist\left[1\right]$, thus at the end of the execution,
    \begin{gather*}
      \mathrm{sc}_{\rounds}\left(\postlist\left[1\right]\right) = \left(k -
      1\right)\left(a + b\right) + 2\left(\gamma a + b\right) + \left(k -
      1\right)\left(\left(\gamma - \epsilon\right)\rvp a + b\right) \enspace, \\
      \mathrm{sc}_{\rounds}\left(\postlist\left[2\right]\right) = \left(k -
      1\right)\left(a + b\right) + \left(k + 1\right)\left(\gamma \rvp a +
      b\right) \enspace.
    \end{gather*}
    Given the bound on $\epsilon$, it is
    $\mathrm{sc}_{\rounds}\left(\postlist\left[1\right]\right) >
    \mathrm{sc}_{\rounds}\left(\postlist\left[2\right]\right)$, thus
    $\mathrm{Ideal}^1\left(\postlist_{\rounds}\right)$ does not hold.

    Second, we consider the case when $\playerlen$ is odd: $\exists k \in
    \mathbb{N}: \playerlen = 2k + 1$. Let $0 < \gamma < 1, 0 <
    \epsilon < \gamma \frac{1 - \rvp}{k\rvp}$,
    \begin{gather*}
      \weakpost = \left(\underbrace{1, \dots, 1}_{k}, \underbrace{\gamma -
      \epsilon, \dots, \gamma - \epsilon}_{k}, \gamma\right) \enspace, \\
      \strongpost = \left(\underbrace{\gamma, \dots, \gamma}_{k}, \underbrace{1,
      \dots, 1}_{k}, \gamma\right) \enspace, \\
      \postlist = \left[\weakpost, \strongpost, \underbrace{\nullpost, \dots,
      \nullpost}_{\postlen - 2}\right] \enspace.
    \end{gather*}
    We first observe that
    \begin{gather*}
      \forall j \in \left\{3, \dots, \postlen\right\},
      \idsc{\postlist\left[j\right]} = 0 < \\
      < \idsc{\postlist\left[1\right]} = k + k\left(\gamma - \epsilon\right) +
      \gamma = \\
      = k + \left(k + 1\right)\gamma -k\epsilon < k + \left(k + 1\right)\gamma =
      \idsc{\postlist\left[2\right]} \enspace,
    \end{gather*}
    thus the strong post has strictly the highest ideal score of all posts and
    as a result, $\forall \postlist'$ that contains the same posts as
    $\postlist$ and $\textsc{Ideal}^1\left(\postlist'\right)$ holds, it is
    $\postlist'\left[1\right] = \postlist\left[2\right]$.

    The first $k$ and the last player vote first for $\postlist\left[1\right]$
    and then for $\postlist\left[2\right]$, whereas players $k + 1, \dots, 2k$
    vote first for $\postlist\left[2\right]$ and then for
    $\postlist\left[1\right]$, thus at the end of the execution,
    \begin{gather*}
      \mathrm{sc}_{\rounds}\left(\postlist\left[1\right]\right) = k\left(a +
      b\right) + \gamma a + b + k\left(\left(\gamma - \epsilon\right)\rvp a +
      b\right) \enspace, \\
      \mathrm{sc}_{\rounds}\left(\postlist\left[2\right]\right) = k\left(a +
      b\right) + \left(k + 1\right)\left(\gamma \rvp a + b\right) \enspace.
    \end{gather*}
    Given the bound on $\epsilon$, it is
    $\mathrm{sc}_{\rounds}\left(\postlist\left[1\right]\right) >
    \mathrm{sc}_{\rounds}\left(\postlist\left[2\right]\right)$, thus
    $\mathrm{Ideal}^1\left(\postlist_{\rounds}\right)$ does not hold.

    Last but not least, we consider the case when $\playerlen = 2$. In this
    case, let $0 < \gamma < 1$ and
    \begin{equation*}
      \postlist = \left[\left(1, 0\right), \left(\gamma, 1 - \gamma\frac{1 +
      \rvp}{2}\right), \underbrace{\nullpost, \dots, \nullpost}_{\postlen -
      2}\right] \enspace.
    \end{equation*}
    It is $\forall j \in \left\{3, \dots, \postlen\right\},
    \idsc{\postlist\left[j\right]} = 0 < \idsc{\postlist\left[1\right]} = 1
    \overset{\rvp < 1}{<} 1 + \gamma\frac{1 - \rvp}{2} = \gamma + 1 -
    \gamma\frac{1 +\rvp}{2} = \idsc{\postlist\left[2\right]}$, thus
    $\postlist\left[2\right]$ has strictly the highest ideal score of all posts
    and as a result, $\forall \postlist'$ that contains the same posts as
    $\postlist$ and $\textsc{Ideal}^1\left(\postlist'\right)$ holds, it is
    $\postlist'\left[1\right] = \postlist\left[2\right]$.

    On the other hand,
    $\mathrm{sc}_{\rounds}\left(\postlist\left[1\right]\right) = a + 2b > \gamma
    \rvp a + b + \left(1 - \gamma \frac{1 + \rvp}{2}\right) a + b =
    \mathrm{sc}_{\rounds}\left(\postlist\left[2\right]\right)$, thus
    $\mathrm{Ideal}^1\left(\postlist_{\rounds}\right)$ does not hold.
  \end{itemize}
\end{proof}

  \newpage
\section{Steem post voting system procedures}
  \label{appendix:procs}
  \begin{algorithm}[H]
  \caption{$\textsc{Init}\left( \attspan, a, b, \regen, \rounds,
  \stpowvec\right)$}
  \label{alg:steem:init}
  \begin{algorithmic}[1]
    \State Store input parameters as constants
    \State $\round \gets 1$
    \State $\mathrm{lastVoted} \gets \left(0, \dots, 0\right) \in
    \left(\mathbb{N}^*\right)^\playerlen$
    \State $\votpowvec \gets \left(1, \dots, 1\right) \in \left[0,
    1\right]^\playerlen$
    \label{alg:steem:init:vp}
    \State $\mathrm{scores} \gets \left(0, \dots, 0\right) \in
    \left(\mathbb{R}^{+}\right)^\postlen$
  \end{algorithmic}
\end{algorithm}

  \begin{algorithm}[H]
  \caption{\textsc{Aux}}
  \label{alg:steem:aux}
  \begin{algorithmic}[1]
    \State \Return $\left(\attspan, a, b, \round, \regen, \rounds,
    \stpowvec\right)$
  \end{algorithmic}
\end{algorithm}

  \begin{algorithm}[H]
  \caption{$\textsc{HandleVote}\left(\mathrm{ballot},
  \player_{\pid}\right)$}
  \label{alg:steem:handlevote}
  \begin{algorithmic}[1]
    \If{$\mathrm{lastVoted}_{\pid} \neq \round$} \Comment{One vote per
    player per round}
      \State $\votpowvec_{\pid, r} \gets \votpowvec_{\pid}$ \Comment{For
      proofs}
      \State $\votpowvec_{\pid} \gets \max{\lbrace \votpowvec_{\pid} + \regen, 1
      \rbrace}$
      \State $\votpowvecreg_{\pid, r} \gets \votpowvec_{\pid}$ \Comment{For
      proofs}
      \label{alg:steem:handlevote:regen}
      \If{$\mathrm{ballot} \neq \mathbf{null}$}
        \State Parse ballot as $\left(\post, \mathrm{weight}\right)$
        \State $\mathrm{cost} \gets a \cdot \votpowvec_{\pid} \cdot
        \mathrm{weight} + b$
        \label{alg:steem:handlevote:cost:start}
        \If{$\votpowvec_{\pid} - \mathrm{cost} \geq 0$}
          \State $\mathrm{score} \gets \mathrm{cost} \cdot \stpowvec_{\pid}$
          \State $\votpowvec_{\pid} \gets \votpowvec_{\pid} - \mathrm{cost}$
        \Else
          \State $\mathrm{score} \gets \votpowvec_{\pid} \cdot
          \stpowvec_{\pid}$
          \State $\votpowvec_{\pid} \gets 0$
        \EndIf
        \label{alg:steem:handlevote:cost:end}
        \State $\mathrm{scores}_\post \gets \mathrm{scores}_\post +
        \mathrm{score}$
      \EndIf
      \State $\mathrm{lastVoted}_{\pid} \gets \round$
    \EndIf
    \If{$\forall i \in \left[N\right], \mathrm{lastVoted}_i = \round$}
    \Comment{round over}
      \State $\postlist \gets \textsc{Order}\left(\postlist,
      \mathrm{scores}\right)$ \Comment{order posts by votes}
      \label{alg:steem:handlevote:order}
      \State $\postlist_r \gets \postlist$ \Comment{For proofs}
      \State $\round \gets \round + 1$
    \EndIf \Comment{TODO: count rounds? simplify with set of voted and check of
    length?}
  \end{algorithmic}
\end{algorithm}

  \begin{algorithm}[H]
  \caption{$\textsc{Vote}\left(\postlist, \mathrm{aux}\right)$}
  \label{alg:steem:vote}
  \begin{algorithmic}[1]
    \State Store aux contents as constants
    \State $\mathrm{voteRounds} \gets \textsc{VoteRounds}\left(\rounds,
    |\postlist|\right)$
    \If{$\textsc{VoteThisRound}\left(\round, |\postlist|\right) = \mathrm{yes}$}
      \State $\mathrm{top} \gets \textsc{ChooseTopPosts}\left(\attspan,
      \postlist, \mathrm{votedPosts}\right)$
      \State $\left(i, l\right) \gets \argmax\limits_{\left(i,
      l\right) \in \mathrm{top}}{\lbrace l_{\pid} \rbrace}[1]$
      \State $\mathrm{votedPosts} \gets \mathrm{votedPosts} \: \cup \left(i,
      l\right)$
      \State \Return $\left(\left(i, l\right), l_{\pid}\right)$
    \Else
      \State \Return \textbf{null}
    \EndIf
    \State
    \Function{ChooseTopPosts}{$\attspan, \postlist, \mathrm{votedPosts}$}
      \State $\result \gets \emptyset$
      \State $\mathrm{idx} \gets 1$
      \While{$|\result| < \attspan \And \mathrm{idx} \leq |\postlist|$}
        \If{$\postlist\left[\mathrm{idx}\right] \notin \mathrm{votedPosts}$}
        \Comment{One vote per post per player}
          \State $\result \gets \result \cup \lbrace
          \postlist\left[\mathrm{idx}\right] \rbrace$
        \EndIf
        \State $\mathrm{idx} \gets \mathrm{idx} + 1$
      \EndWhile
      \State \Return $\result$
    \EndFunction
    \State
    \Function{VoteThisRound}{$\round, \postlen$}
    \label{alg:steem:vote:votethisround}
      \If{$\rounds < \postlen$}
        \State \Return yes
      \ElsIf{$\round \in \mathrm{voteRounds}$}
        \State \Return yes
      \Else
        \State \Return no
      \EndIf
    \EndFunction
    \State
    \Function{VoteRounds}{$\rounds, \postlen$}
    \label{alg:steem:vote:voterounds:start}
      \State $\mathrm{voteRounds} \gets \emptyset$
      \For{$i = 1$ to $\postlen$}
        \State $\mathrm{voteRounds} \gets \mathrm{voteRounds} \cup \left\{1 +
        \floor*{\left(i - 1\right)\frac{\rounds - 1}{\postlen - 1}}\right\}$
      \EndFor
      \State \Return voteRounds
    \EndFunction
    \label{alg:steem:vote:voterounds:end}
  \end{algorithmic}
\end{algorithm}

\end{subappendices}

\bibliography{latex/references}
\end{document}